\documentclass{emulateapj}
\usepackage{graphicx}
\usepackage{amssymb,amsfonts,amsmath,amstext,amsgen,amsopn,amsxtra,indentfirst,times}

\begin{document}

\title{\texttt{HELIOS-K}: An Ultrafast, Open-source Opacity Calculator for Radiative Transfer}
\author{Simon L. Grimm\altaffilmark{1} \& Kevin Heng\altaffilmark{2}}
\altaffiltext{1}{University of Z\"{u}rich, Institute for Computational Science, Winterthurerstrasse 190, CH-8057, Z\"{u}rich, Switzerland.  Email: sigrimm@physik.uzh.ch}
\altaffiltext{2}{University of Bern, Physics Institute, Center for Space and Habitability, Sidlerstrasse 5, CH-3012, Bern, Switzerland.  Email: kevin.heng@csh.unibe.ch}

\begin{abstract}
We present an ultrafast opacity calculator that we name \texttt{HELIOS-K}.  It takes a line list as an input, computes the shape of each spectral line and provides an option for grouping an enormous number of lines into a manageable number of bins.  We implement a combination of Algorithm 916 and Gauss-Hermite quadrature to compute the Voigt profile, write the code in \texttt{CUDA} and optimise the computation for graphics processing units (GPUs).  We restate the theory of the k-distribution method and use it to reduce $\sim 10^5$--$10^8$ lines to $\sim 10$--$10^4$ wavenumber bins, which may then be used for radiative transfer, atmospheric retrieval and general circulation models.   The choice of line-wing cutoff for the Voigt profile is a significant source of error and affects the value of the computed flux by $\sim 10\%$.  This is an outstanding physical (rather than computational) problem, due to our incomplete knowledge of pressure broadening of spectral lines in the far line wings.  We emphasize that this problem remains regardless of whether one performs line-by-line calculations or uses the k-distribution method and affects all calculations of exoplanetary atmospheres requiring the use of wavelength-dependent opacities.  We elucidate the correlated-k approximation and demonstrate that it applies equally to inhomogeneous atmospheres with a single atomic/molecular species or homogeneous atmospheres with multiple species.  Using a NVIDIA K20 GPU, \texttt{HELIOS-K} is capable of computing an opacity function with $\sim 10^5$ spectral lines in $\sim 1$ second and is publicly available as part of the Exoclimes Simulation Platform (ESP; \texttt{www.exoclime.org}).
\end{abstract}


\section{Introduction}

\subsection{The million- to billion-line radiative transfer challenge}

Measuring the spectra of exoplanetary atmospheres gives us a window into their thermal structure and chemical compositions \citep{brown01,burrows01,char09,sd10,madhu14,hs15}.  A crucial bridge between observation and inference is the use of theoretical models of atmospheric radiation, both in the form of ``forward models" that adopt a set of fixed assumptions (e.g., solar composition) and retrieval models that attempt to invert for various properties from the data.  In both families of models, one needs to compute synthetic spectra, which in turn requires the computation of the opacity function of the atmosphere.

To achieve a high degree of accuracy, it is desirable to perform ``line-by-line" calculations, where every spectral line in the range of wavelengths considered, for a given molecule (e.g., water), is directly included either in the process of solving for radiative equilibrium (in forward models) or a multi-parameter search for an optimal solution based on a comparison to data (in retrieval models).  Such an approach may be readily adopted at low temperatures, but at the high temperatures ($\sim 800$---3000 K) of the exoplanetary atmospheres currently amenable to characterisation by astronomy, it becomes infeasible as the number of spectral lines involved increases by orders of magnitude.  For example, the HITRAN database lists $\sim 10^5$ lines for the water molecule, but is only valid up till temperatures of about 800 K.  At higher temperatures, millions of weak lines become important and the total number of lines involved increases to $\sim 10^8$; the HITEMP database needs to be used instead.  Line-by-line calculations become expensive or even prohibitive as one attempts to explore the broad parameter space occupied by exoplanetary atmospheres.  Furthermore, in studies where line-by-line calculations are claimed, it is not always clear that sufficient resolution has been devoted to computing the $\gtrsim 10^8$ lines of the opacity function for hot exoplanetary atmospheres.  As different combinations of molecules, temperature and pressure are considered, the problem becomes computationally intractable.

\subsection{The method of k-distribution tables}

\begin{figure}
\begin{center}
\vspace{-0.1in}
\includegraphics[width=\columnwidth]{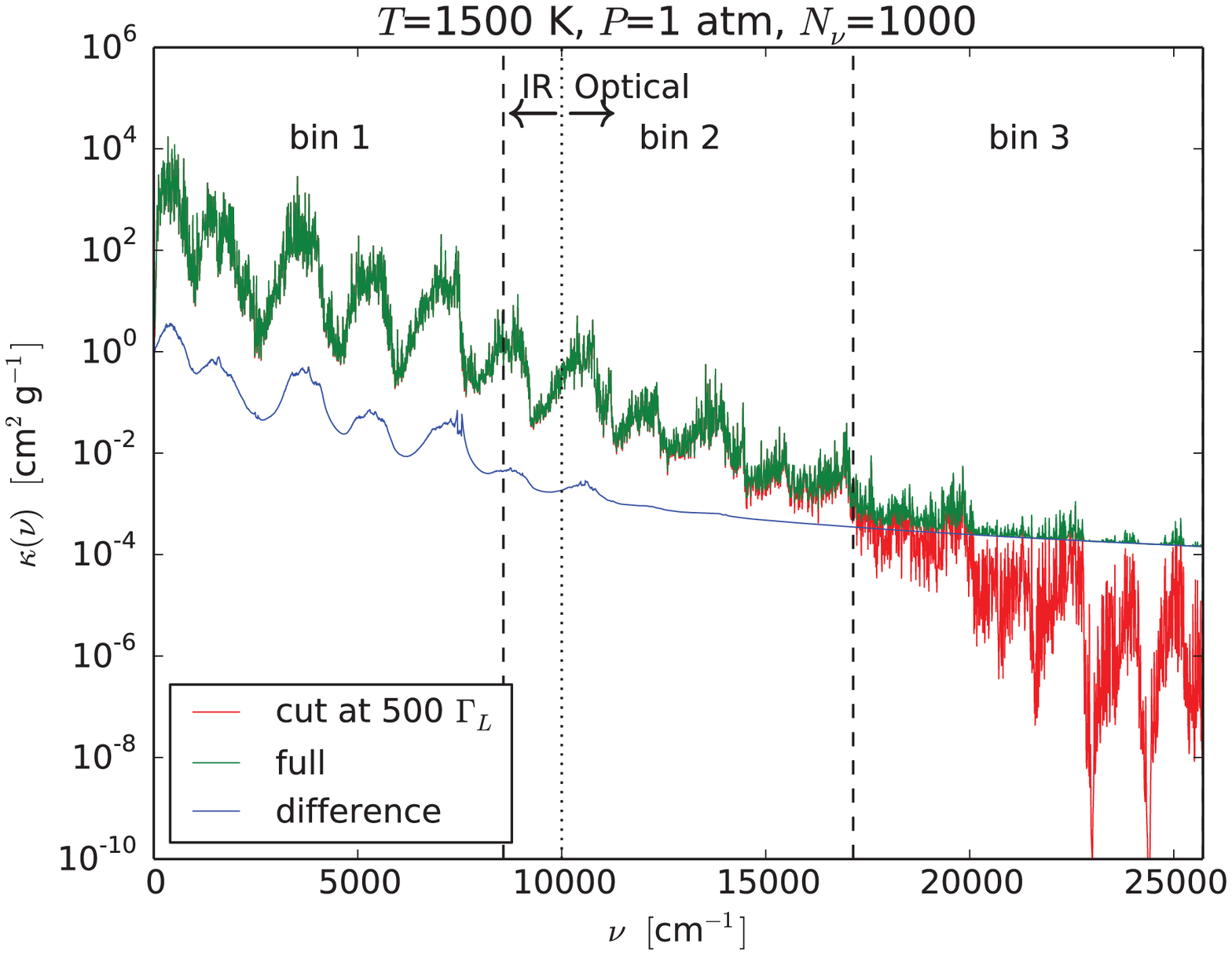}
\includegraphics[width=\columnwidth]{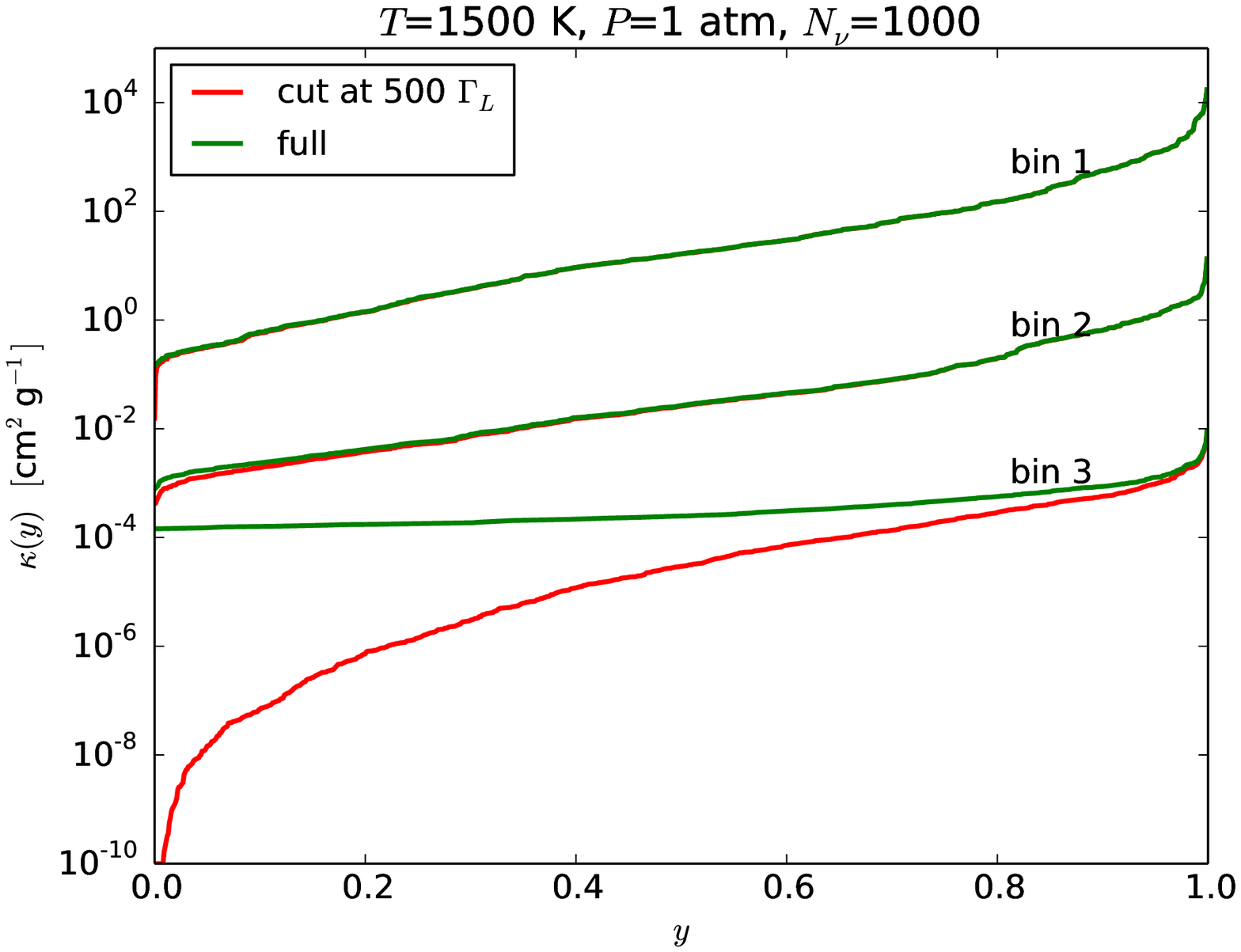}
\includegraphics[width=\columnwidth]{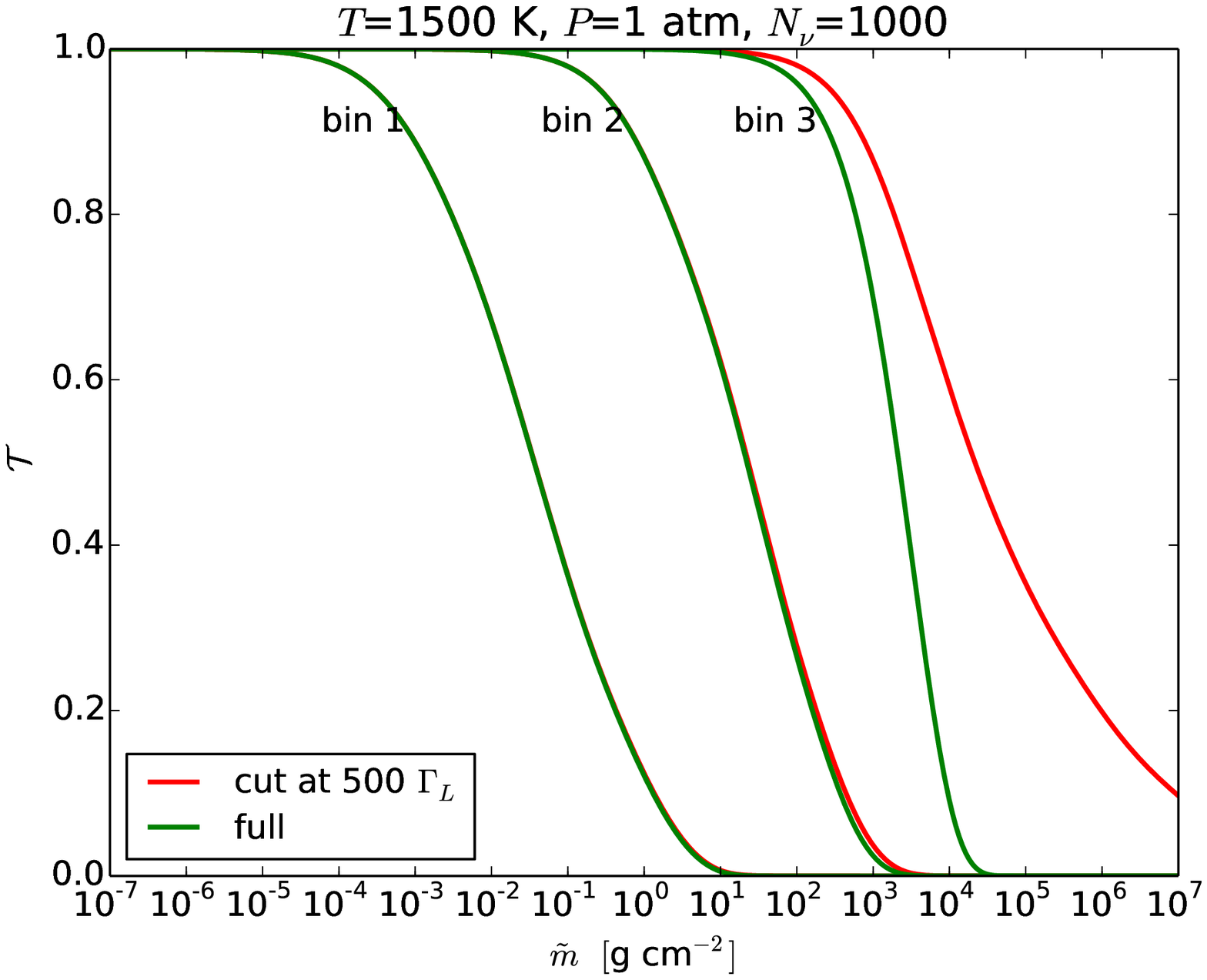}
\end{center}
\vspace{-0.2in}
\caption{To highlight the salient features explored in this study, we divide the opacity function of the water molecule, as provided by the HITEMP database, into three different regions, which we term ``bin 1", ``bin 2" and ``bin 3" in this montage of figures.  These bins cover the infrared, infrared-optical transitional and optical range of wavelengths.  Dividing the opacity function into more bins does not alter our qualitative conclusions.  As an illustration, we adopt numbers representative of hot exoplanetary atmospheres: $T=1500$ K and $P = 1$ atm $= 0.98692$ bar.  Top panel: opacity function using spectroscopic quantities from the HITEMP database.  Shown are calculations using the full Voigt function and with an ad hoc line-wing cutoff of $500 \Gamma_{\rm L}$.  Middle panel: k-distribution functions for the three wavenumber different regions of the opacity function.  Bottom panel: transmission function corresponding to the three wavenumber regions, both with and without the Voigt line-wing cutoff.}
\label{fig:main}
\end{figure}

In the Earth and planetary sciences, a well-worn strategy for dealing with an enormous number of lines is the method of ``k-distribution tables"\footnote{We regard this term as being a synonym, since we will always denote opacities by $\kappa$ and not ``k", following the convention in some parts of the astrophysics literature.  However, to preserve tradition we will retain the name ``k-distribution".} \citep{gy89,lo91,fu92}.  The essence of the method is to perform Lebesgue, instead of Riemann, integration \citep{pierrehumbert}, when integrating over the opacity function of the atmosphere to determine if it is transparent or opaque within a given spectral window.  Instead of integrating over the opacity function itself, which is computationally unwieldy as it is hardly a smooth and predictable function, one recasts it into its cumulative counterpart---a smooth, monotonically increasing and computationally pleasing function.  This cumulative function may then be used to compute the transmission function: it is the fraction of radiation passing from one layer of the atmosphere to the next within a given spectral window.  Figure \ref{fig:main} shows an example of this process.

The cumulative counterpart of the opacity function is known as the ``k-distribution function".  The term ``k-distribution table" is commonly used, because this cumulative function may be tabulated beforehand and then used to perform integrations in forward models of radiative transfer (e.g., \citealt{marley96,burrows97,fortney10}), retrieval models (e.g., \citealt{lee12}) and three-dimensional simulations of atmospheric circulation (e.g., \citealt{showman09}).

Nevertheless, several physical and computational issues remain either unelucidated or poorly elucidated within the literature, which provide the motivation behind the current study.  Our main, physical conclusion is that \textit{physical} (and not computational) uncertainties associated with the wings of spectral lines dominate the error budget.  Our technical contribution is an ultrafast, open-source computer code to compute the opacity function using modern computing methods and architectures.  

\section{Method}

\subsection{Theory of k-distribution method versus correlated-k approximation}

\subsubsection{Restatement of basic theory of k-distributions}

Consider an arbitrary function $f(x)$, where $x$ is the wavenumber\footnote{We use wavenumber, instead of wavelength, because it is the preferred choice of spectroscopic databases like HITRAN and HITEMP and spectral lines are more evenly spaced across wavenumber (or frequency) than wavelength.} normalized by the entire range considered.  We wish to evaluate the integral over the range $x_{\rm min} \le x \le x_{\rm max}$,
\begin{equation}
I = \int^{x_{\rm max}}_{x_{\rm min}} f\left(x\right) ~dx.
\end{equation}
Imagine that $f(x)$ may be recast as $f(y)$ such that the quantity $y$ is the fractional area under the curve that satisfies $f(x) \le f_0$, where $f_0$ is an arbitrary value of the function.  Then, the same integral may be evaluated as
\begin{equation}
I = \int^1_0 f\left(y\right) ~dy.
\end{equation}
Practically all of the functions we encounter in astrophysics may be integrated using this alternative expression.

More generally, we have
\begin{equation}
\int F ~dx = \int {\cal H} ~dF,
\end{equation}
where ${\cal H}$ is the fractional cumulative distribution function of another arbitrary function, $F(f(x))$, that satisfies $F \le F_0$ and $F_0$ is an arbitrary value of $F$.  In other words, ${\cal H}$ gives the fractional area under the curve corresponding to $F \le F_0$.

We make these concepts less abstract by applying them to an atmosphere.  In the simplest case, suppose that $F = f = \kappa$, where $\kappa$ is the opacity function (with units of cm$^2$ g$^{-1}$).  It follows that $\int {\cal H} ~dF = \int {\cal H} \kappa ~dx = \int \kappa ~dy$.  The quantity $y = \int^x_0 {\cal H} ~dx$ is the cumulative sum of intervals.  As expected, one gets the same answer whether one evaluates $\int \kappa ~dx$ or $\int \kappa ~dy$.  A more useful example considers
\begin{equation}
f = \kappa, ~F = \exp{\left( - \kappa \tilde{m} \right)},
\end{equation}
where $\tilde{m}$ is the column mass, since the transmission function,
\begin{equation}
{\cal T} = \int^\infty_0 F ~dx = \int^1_0 F ~dy,
\label{eq:transmission}
\end{equation}
is a quantity that is indispensible for computing synthetic spectra (e.g., \citealt{hml14}).  The transmission function is commonly integrated over some wavelength range and is the degree or transparency (or opaqueness) of this spectral window.  For example, in a purely absorbing atmosphere the flux passing from one layer to another is given by $F_{\rm layer} = F_{\rm previous} {\cal T} + \pi B \left( 1 - {\cal T} \right)$, where $F_{\rm previous}$ is the flux from the previous layer and $B$ is the Planck function (e.g., \citealt{hml14}).  The second equality in equation (\ref{eq:transmission}) obtains from expressing the cumulative sum of intervals as
\begin{equation}
y = \int^\kappa_0 {\cal H} \tilde{m} ~d\kappa.
\end{equation}
We will refer to $\kappa(y)$ as the ``k-distribution function".

\subsubsection{Correlated-k approximation}

The k-distribution method is exact for a homogeneous atmosphere, which almost never happens in practice.  For an inhomogeneous atmosphere, the opacity changes with the temperature and pressure and we have
\begin{equation}
\begin{split}
{\cal T} &= \int^\infty_0 \exp{\left[ -\int \kappa\left(x\right) ~d\tilde{m} \right]} ~dx \\
&\ne \int^1_0 \exp{\left[ -\int \kappa\left(y\right) ~d\tilde{m} \right]} ~dy.
\end{split}
\label{eq:corr_k}
\end{equation}

That the k-distribution method cannot be used for an inhomogeneous atmosphere may be illustrated using the example of a two-layered atmosphere.  Each layer has its own opacity function and column mass (subscripted by ``1" and ``2") and the transmission function is
\begin{equation}
\begin{split}
{\cal T} =& \int^1_0 \exp{\left[ - \kappa_1\left(y_1\right) \tilde{m}_1 -\kappa_2\left(y_1\right) \tilde{m}_2 \right]} ~dy_1 \\
&+\int^1_0 \exp{\left[ - \kappa_1\left(y_2\right) \tilde{m}_1 -\kappa_2\left(y_2\right) \tilde{m}_2 \right]} ~dy_2 \\
\ne& 2 \int^1_0 \exp{\left[ - \kappa_1\left(y\right) \tilde{m}_1 -\kappa_2\left(y\right) \tilde{m}_2 \right]} ~dy.
\end{split}
\label{eq:transfunc}
\end{equation}
That there are two integrals originates from having $F = \exp{(-\kappa_1 \tilde{m}_1 - \kappa_2 \tilde{m}_2)}$ and
\begin{equation}
dF = -F \left( \tilde{m}_1 d\kappa_1 + \tilde{m}_2 d\kappa_2 \right).
\end{equation}
Also, we have
\begin{equation}
dy_1 = {\cal H} \tilde{m}_1 ~d\kappa_1~, dy_2 = {\cal H} \tilde{m}_2 ~d\kappa_2.
\end{equation}

The non-equality in equation (\ref{eq:transfunc}) derives from the fact that \textit{even identical ranges of values in $y_1$ and $y_2$ generally correspond to different ranges of wavenumbers}.  For example, $\kappa_1(y_1)$ and $\kappa_2(y_2)$ are cumulative functions constructed from their own cumulative sum of intervals.  By contrast, $\kappa_1(y_2)$ and $\kappa_2(y_1)$ are cumulative functions constructed from the cumulative sum of intervals of their counterparts, meaning that the contributions are drawn from different wavenumber intervals even at the same value of the cumulative sum of intervals.  Generally, we expect these four cumulative functions to have different functional forms.  This peculiar property is an unavoidable consequence of working with cumulative functions.

\textit{Physically, in employing the k-distribution method, the price being paid is that the wavenumber information has been scrambled.}  If one \textit{assumes} that $y = y_1 = y_2$, then one is making the ``correlated-k approximation" and the transmission function may then be computed as a single integral across $y$.  It is the assumption that each value of the cumulative opacity function is always drawn from the same wavenumber interval.

The mathematics behind the reasoning is identical in the case of applying the correlated-k approximation to a homogeneous atmosphere with multiple atoms or molecules.  For illustration, consider only two molecules and a single value of the column mass.  Let the mixing ratios (relative abundance by number) of the molecules be $X_1$ and $X_2$.  We then have
\begin{equation}
\begin{split}
{\cal T} =& \int^1_0 \exp{\left[ - X_1 \kappa_1\left(y_1\right) \tilde{m} - X_2 \kappa_2\left(y_1\right) \tilde{m} \right]} ~dy_1 \\
&+\int^1_0 \exp{\left[ - X_1 \kappa_1\left(y_2\right) \tilde{m} - X_2 \kappa_2\left(y_2\right) \tilde{m} \right]} ~dy_2 \\
\ne& 2 \int^1_0 \exp{\left[ - X_1 \kappa_1\left(y\right) \tilde{m} - X_2 \kappa_2\left(y\right) \tilde{m} \right]} ~dy.
\end{split}
\end{equation}
Here, the fact that we have two integrals comes from having $F = \exp{[-( X_1 \kappa_1 + X_2 \kappa_2 ) \tilde{m}]}$ and
\begin{equation}
dF = -F \tilde{m} \left( X_1 d\kappa_1 + X_2 d\kappa_2 \right).
\end{equation}
Also, we have
\begin{equation}
dy_1 = {\cal H} \tilde{m} X_1 ~d\kappa_1~, dy_2 = {\cal H} \tilde{m} X_2 ~d\kappa_2.
\end{equation}
We have intentionally written things out explicitly to illustrate the fact that one can avoid dealing with two integrals if a single, total opacity function is constructed first ($\kappa = X_1 \kappa_1 + X_2 \kappa_2$) \textit{before} its cumulative function is computed.

Again, unless $y_1=y_2$, the two integrals cannot be combined.  Since this reasoning holds for multiple molecules in a homogeneous atmosphere, it must also hold for multiple molecules in an inhomogeneous atmosphere.  We conclude that one needs to first add the opacities of the various molecules in an atmosphere, weighted by their relative abundances, prior to constructing the cumulative function of the opacity.  If one adds the cumulative opacity functions of different molecules, then one is effectively employing the correlated-k approximation.

Both lines of reasoning can be straightforwardly generalised to an inhomogeneous atmosphere containing a single atom or molecule and with $N$ layers, a homogenous atmosphere with $N$ atomic or molecular species, or an inhomogeneous atmosphere with an arbitrary number of layers and species.

A common source of confusion in the literature is the failure to distinguish the method (k-distribution) from the approximation (correlated-k).  For example, the ``correlated-k method" is a misnomer.

\subsection{Implementing the k-distribution method}

Consider equal intervals in $x$ and let the interval be denoted by $\delta x$.  Such a uniform grid in $x$ generally leads to a non-uniform grid in $\kappa(x)$.  Its virtue is that it reduces our problem to one of sorting and ordering, since every value of $\kappa(x)$ is associated with $\delta x$ (and we do not have to keep track of changing values of the interval).  For a fixed value of the opacity ($\kappa_0$), we count the number of points that satisfy $\kappa(x) \le \kappa_0$.  If $N_{\rm x}$ points are counted, then we have
\begin{equation}
y = \frac{N_{\rm x} ~\delta x}{\Delta x},
\end{equation}
where $\Delta x = x_{\rm max} - x_{\rm min}$ is the range of $x$ being considered.  We also have $\delta x = \Delta x / N_\nu$, where $N_\nu$ is the total number of intervals in $x$.  It implies that the interval in $y$ is also equal,
\begin{equation}
\delta y = \frac{\delta x}{\Delta x} = \frac{1}{N_\nu}.
\end{equation}

By running through all possible values of $\kappa_0$, one constructs $\kappa(y)$.  Since $\kappa(y)$ is a monotonic function that is typically smoother than $\kappa(x)$, it may be resampled and defined over a much smaller number of points, $N_{\rm y} \ll N_\nu$.  It is then used to calculate ${\cal T}$ for any value of $\tilde{m}$.

\subsection{Using the HITRAN and HITEMP databases}

The opacity function is a product of two quantities: the integrated line strength ($S$) and the line profile or shape ($\Phi$) \citep{gy89},
\begin{equation}
\kappa = S \Phi.
\end{equation}
The integrated line strength depends only on the temperature ($T$), while $\Phi$ depends on both temperature and pressure ($P$).  Note that some references collectively refer to opacities (with units of cm$^2$ g$^{-1}$), cross sections (with units of cm$^2$) and absorption coefficients (with units of cm$^{-1}$) as ``absorption coefficients" (e.g., Appendix 2 of \citealt{gy89}).  Only when $\kappa$ is an actual opacity is $S = S(T)$ with no dependence on pressure.

By invoking the principle of detailed balance and local thermodynamic equilibrium, one obtains \citep{penner52,rothman96},
\begin{equation}
S = \frac{g_2 A_{21}}{8 \pi c \nu^2 m Q} ~\exp{\left( - \frac{\Delta E}{k_{\rm B} T} \right)} \left[ 1 - \exp{\left( - \frac{h c \nu}{k_{\rm B} T} \right)} \right],
\end{equation}
where $g_2$ is the statistical weight of the upper level (of a given line transition), $A_{21}$ is the Einstein A-coefficient, $c$ is the speed of light, $\nu$ is the wavenumber, $m$ is the mean molecular mass, $Q$ is the partition function, $\Delta E$ is the energy difference associated with the line transition, $k_{\rm B}$ is Boltzmann's constant and $h$ is Planck's constant.  The partition function relates the number density associated with an energy level with the total number density and is a function of $T$.

In practice, a more useful expression for the integrated line strength is \citep{rothman96},
\begin{equation}
\frac{S}{S_0} = \frac{Q_0}{Q} \exp{\left( - \frac{\Delta E}{k_{\rm B} T} + \frac{\Delta E}{k_{\rm B} T_0} \right)} \frac{1 - \exp{\left( - h c \nu/k_{\rm B} T \right)}}{1 - \exp{\left( - h c \nu/k_{\rm B} T_0 \right)}},
\end{equation}
where all of the quantities subscripted with a ``0" are specified at a reference temperature, $T_0$.  The HITRAN \citep{rothman13} and HITEMP \citep{rothman10} databases provide tabulated values of all of the quantities needed to construct $S$ using $T_0 = 296$ K.

The Voigt profile is the convolution of the Lorentz and the Doppler profiles (e.g., \citealt{draine11}),
\begin{equation}
\begin{split}
\Phi &= \left( \frac{\ln2}{\pi} \right)^{1/2} \frac{H_{\rm V}}{\Gamma_{\rm D}},\\
H_{\rm V} &= \frac{a}{\pi} \int^{+\infty}_{-\infty} \frac{\exp{\left( u^{\prime2} \right)}}{\left( u - u^\prime \right)^2 + a^2} ~du^\prime, 
\end{split}
\end{equation}
where $\Gamma_{\rm D} = \nu_0 \sqrt{2 \ln{2} ~k_{\rm B} T / m} / c$ is the half-width at half-maximum of the Doppler profile, $\nu_0$ is the line-center wavenumber, $a = \sqrt{\ln2}\Gamma_{\rm L}/\Gamma_{\rm D}$ is the damping parameter and $u = \sqrt{\ln2} (\nu-\nu_0)/\Gamma_{\rm D}$.  Our definitions for $\Gamma_{\rm D}$, $a$ and $u$ depart slightly from the traditional ones in order to be consistent with \cite{lb07}.  We have included the effects of pressure broadening within our definition of the half-width of the Lorentz profile \citep{mihalas,rothman96},
\begin{equation}
\Gamma_{\rm L} = \frac{A_{21}}{4\pi c} + \left( \frac{T}{T_0} \right)^{-n_{\rm coll}} \left[ \frac{\alpha_{\rm air} \left(P - P_{\rm self} \right)}{P_0} + \frac{\alpha_{\rm self} P_{\rm self}}{P_0} \right],
\end{equation}
where the first term after the equality is typically subdominant.  Pressure broadening is included via an empirical fit \citep{rothman96}, whose fitting parameters ($n_{\rm coll}$, $\alpha_{\rm air}$ and $\alpha_{\rm self}$) are given by HITRAN and HITEMP.  The reference pressure is $P_0 = 1$ atm $= 0.98692$ bar.  The subscripts ``air" and ``self" represent air- and self-broadening, respectively.  For illustration, we assume that they are present in equal proportions ($P_{\rm self} = 0.5 P$).  We also account for a pressure-induced shift ($\delta_{\rm shift}$) of the central wavenumber, 
\begin{equation}
\nu_0 \rightarrow \nu_0 + \frac{\delta_{\rm shift} P }{ P_0 }, 
\end{equation}
where $\delta_{\rm shift}$ is again a tabulated quantity in HITRAN and HITEMP.  The data for $\delta_{\rm shift}$ is usually sparse.

\subsection{Computing the Voigt profile and the line-wing cutoff problem}

\begin{figure*}
\begin{center}
\vspace{-0.1in}
\includegraphics[width=\columnwidth]{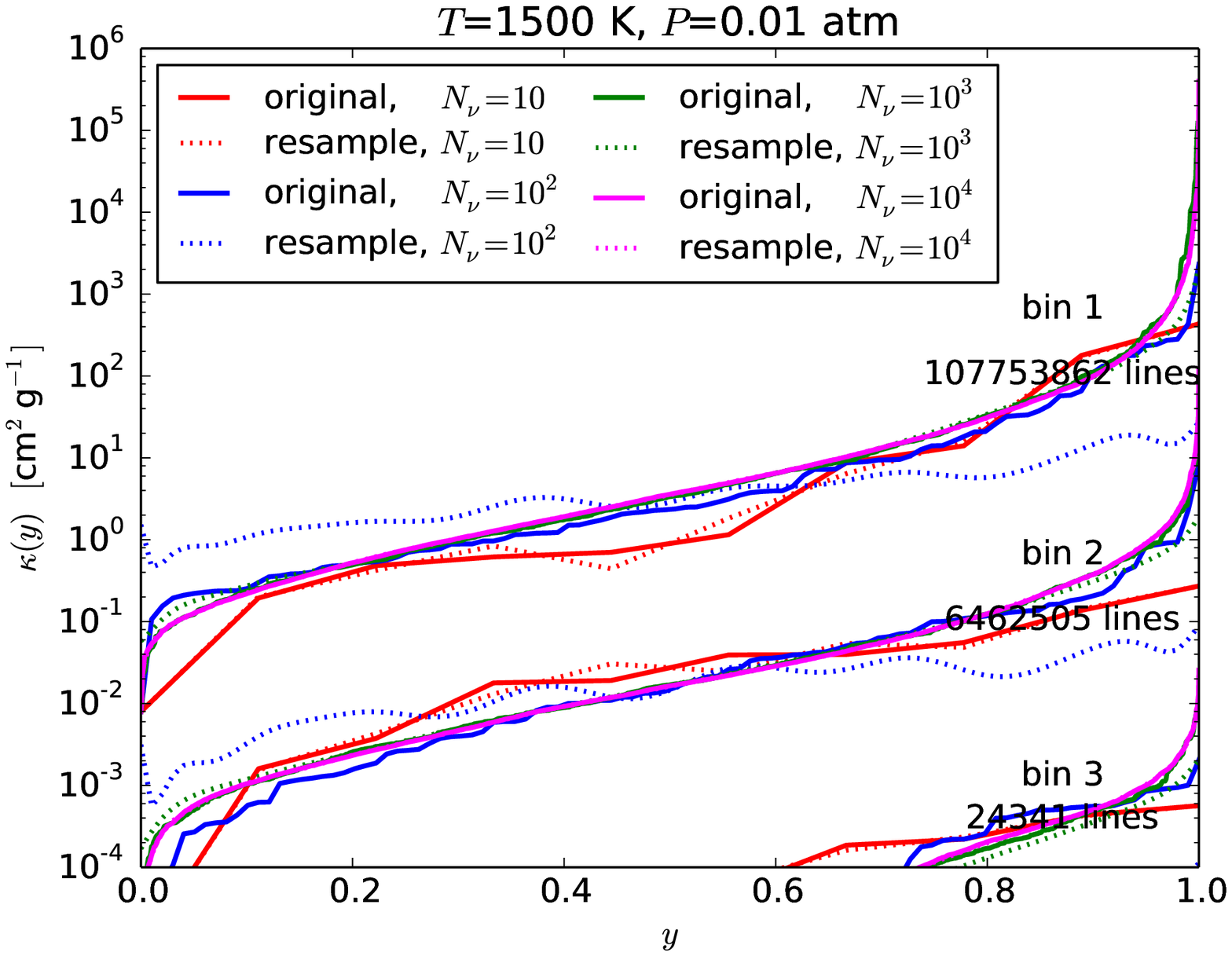}
\includegraphics[width=\columnwidth]{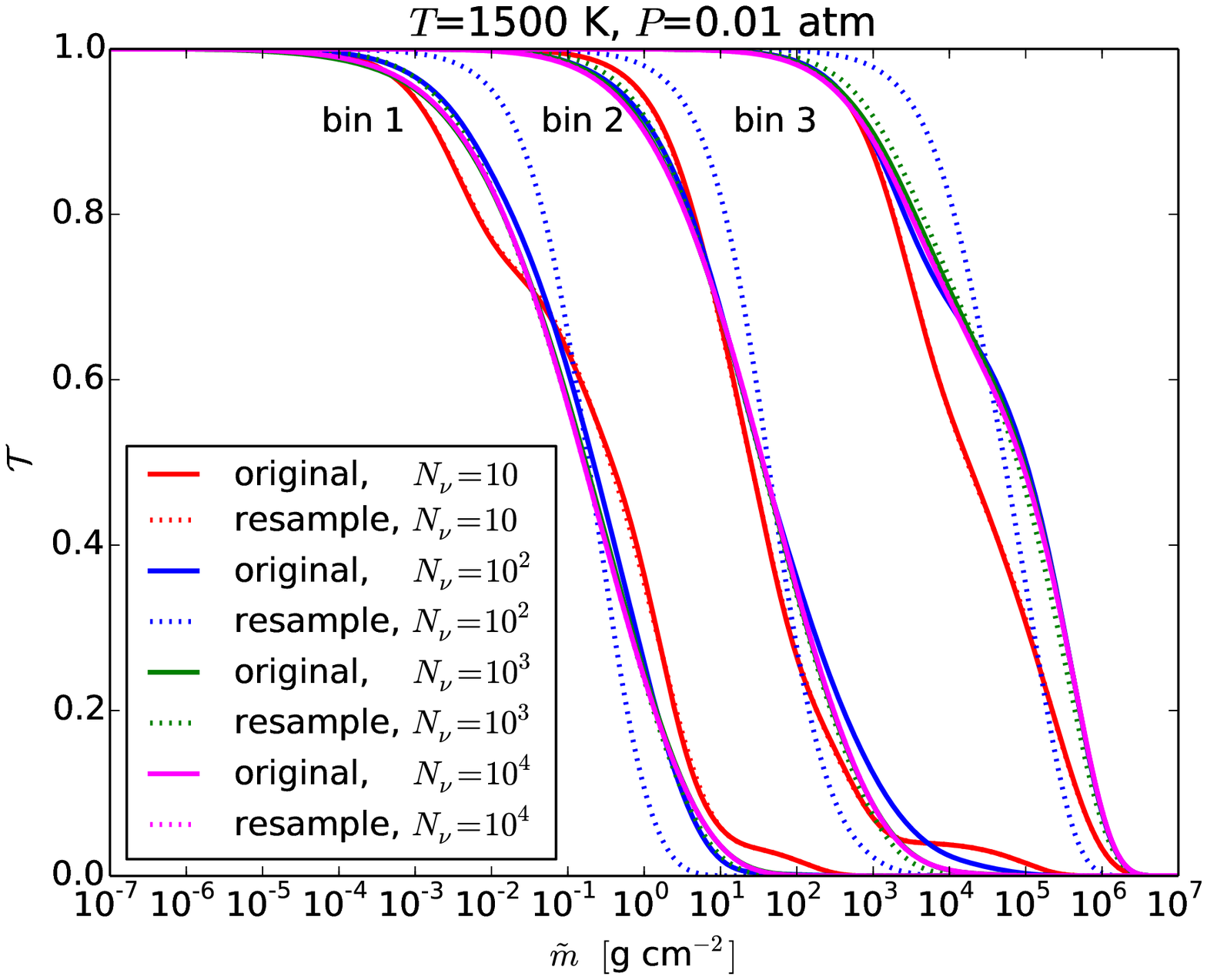}
\includegraphics[width=\columnwidth]{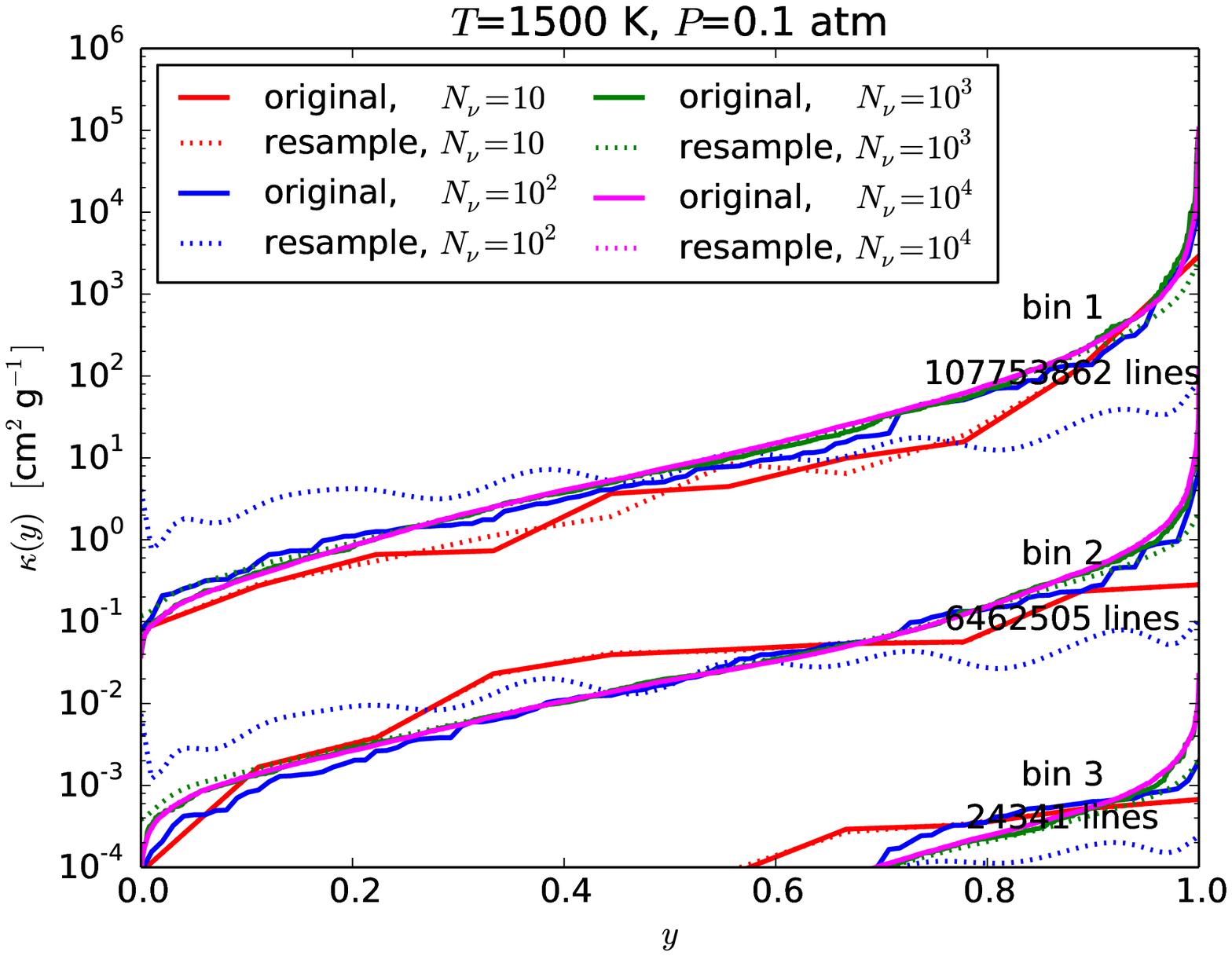}
\includegraphics[width=\columnwidth]{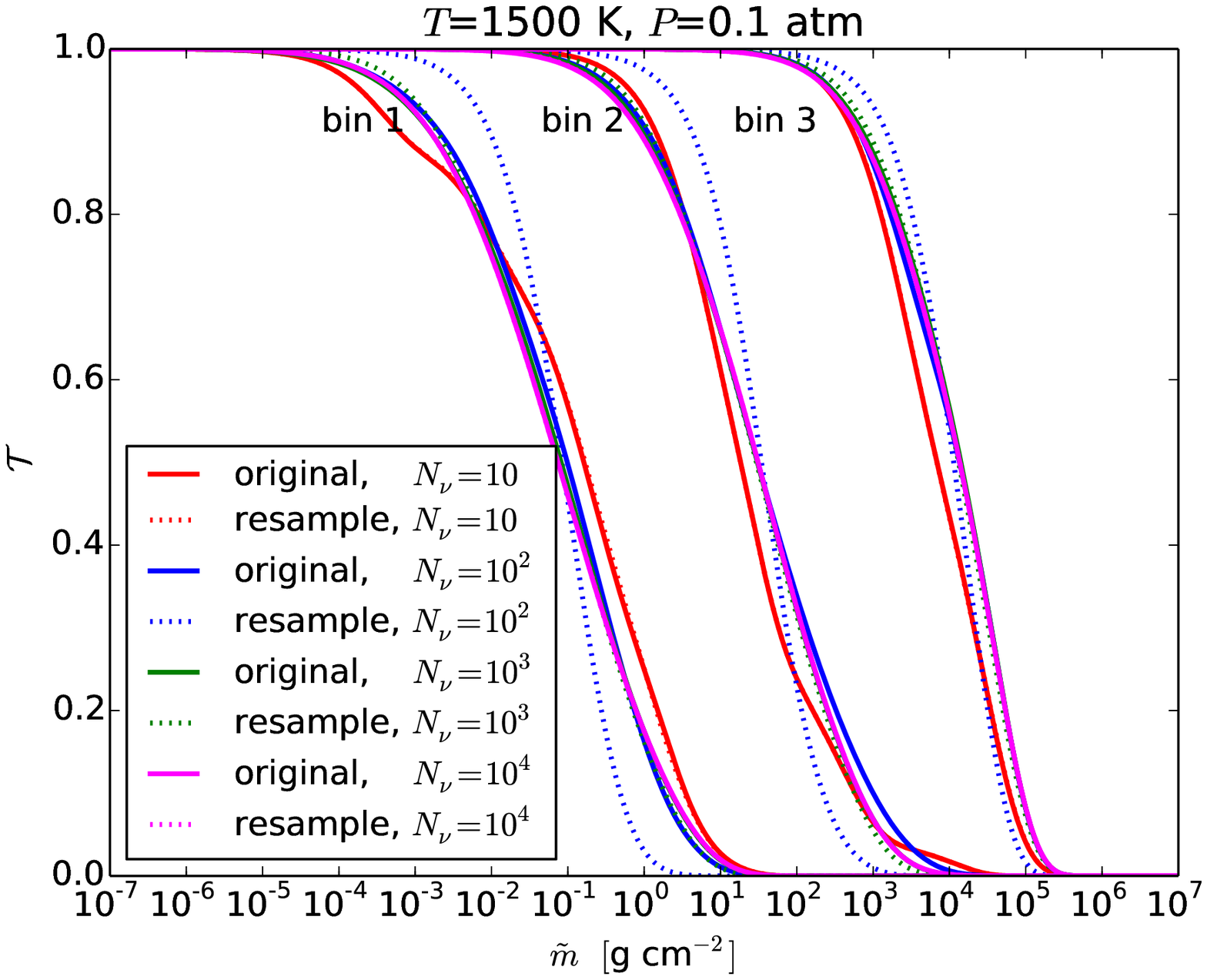}
\includegraphics[width=\columnwidth]{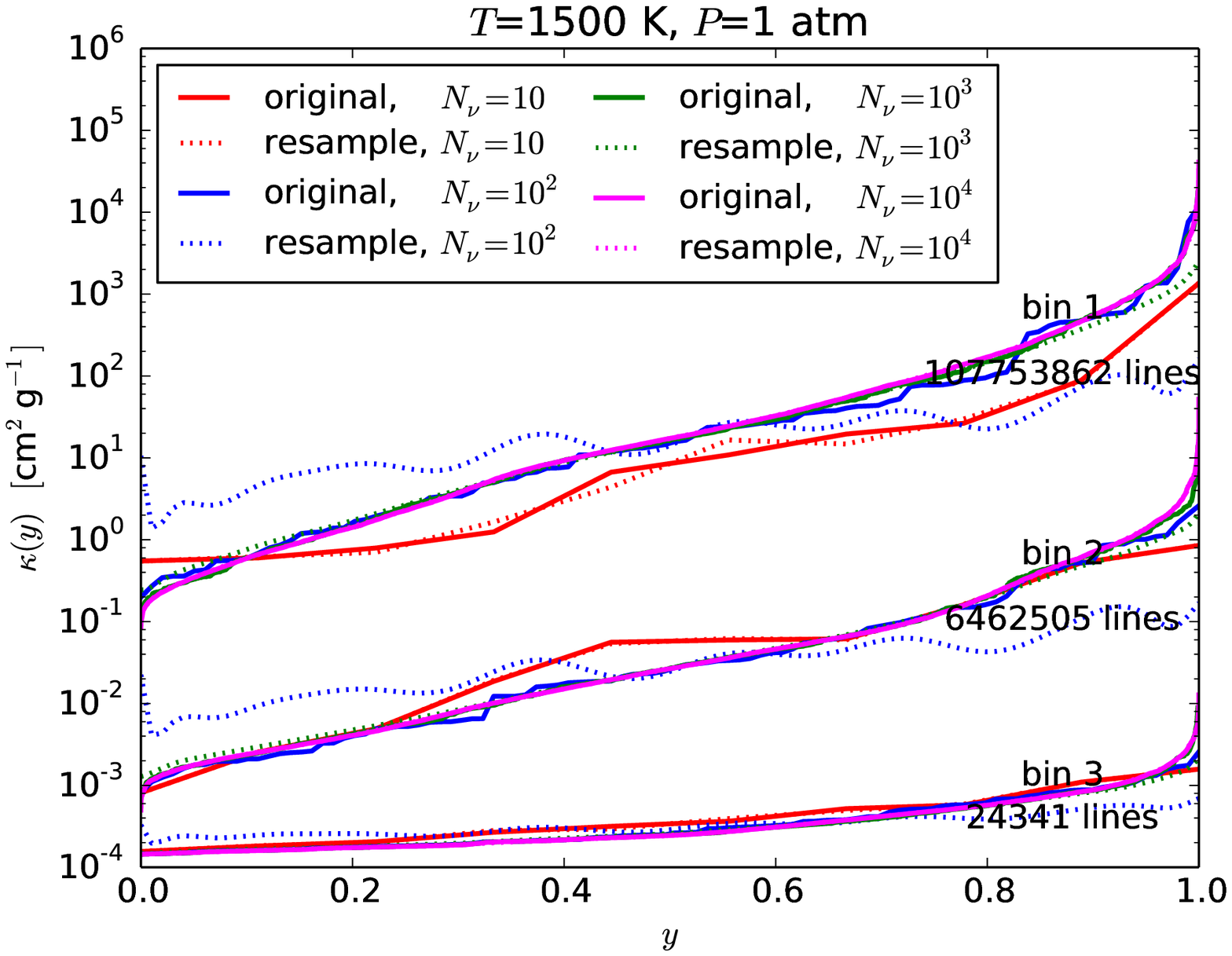}
\includegraphics[width=\columnwidth]{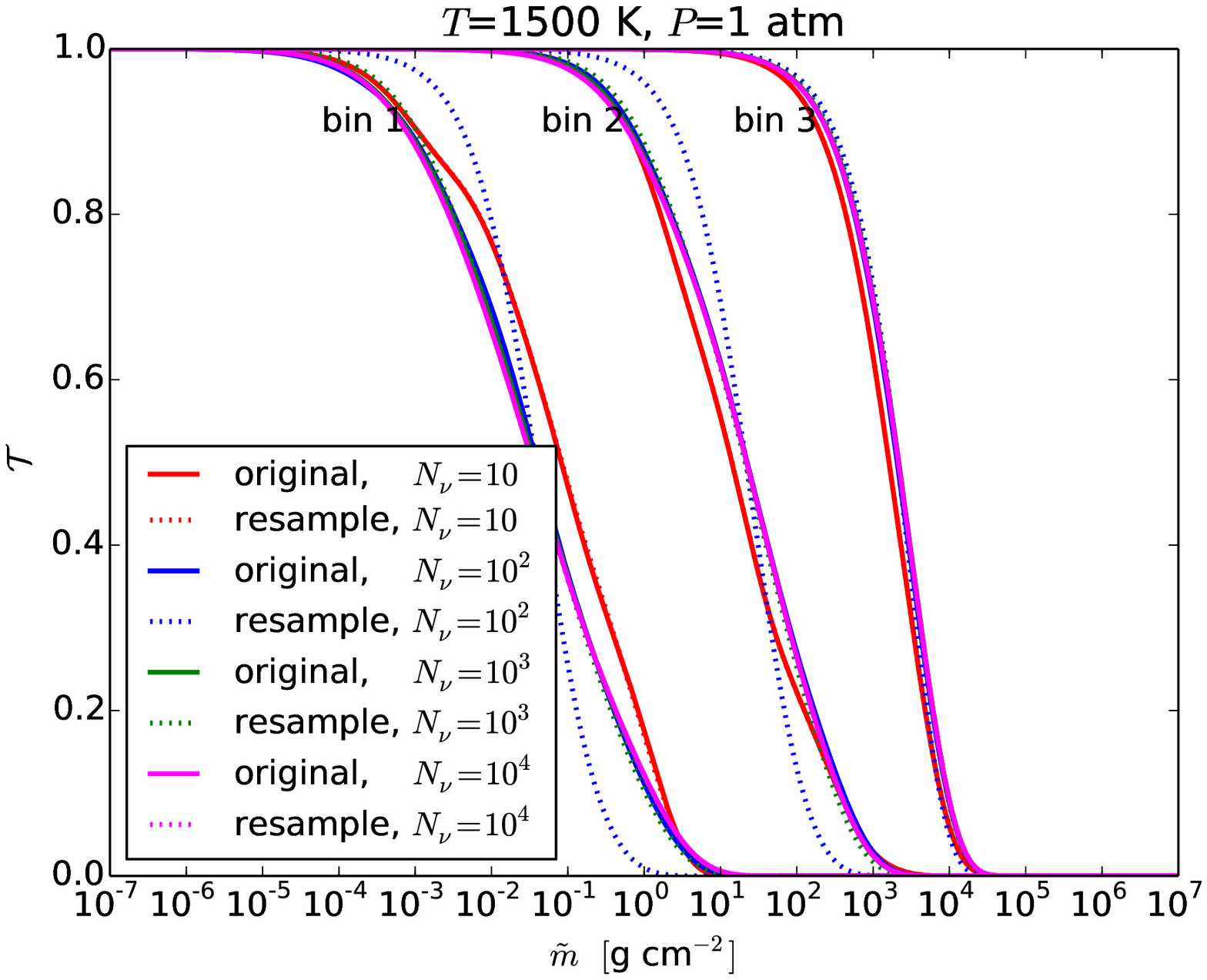}
\end{center}
\vspace{-0.2in}
\caption{Elucidating the effects of resampling and spectral resolution ($N_\nu$).  Left column: k-distribution functions.  Right column: transmission function.  The top, middle and bottom rows are for $P=0.01$, 0.1 and 1 atm, respectively.  For $N_\nu=100$, resampling with 20 Chebyshev coefficients results in discrepancies due to overfitting.}
\label{fig:resampling}
\end{figure*}

There are two challenges associated with the Voigt profile.  The first challenge is computational: it is difficult to evaluate efficiently as it is an indefinite integral.  Furthermore, we have to compute the Voigt profile multiple times for every line and there is an enormous number of lines.  To this end, we implement Algorithm 916, which was originally written for \texttt{MATLAB} \citep{za12}.  The essence of the algorithm is to first recast $H_{\rm V}$ as the real part of the (complex) Faddeeva function and proceed to express it in terms of cosines, sines, a scaled complementary error function and several series expansions, as stated in equations (13), (15), (16) and (17) of \cite{za12}.  The exponential terms in the series expansions are the bottleneck in terms of computational cost; \cite{za12} optimise this process by combining the three series evaluations within a single loop.

It turns out that Algorithm 916 is efficient only for small values of $a$ and $u$.  For $a^2 + u^2 \ge 100$, we implement third-order Gauss-Hermite quadrature to compute $H_{\rm V}$ as stated in equation (8) of \cite{lb07}.  For $a^2 + u^2 \ge 10^6$, we switch from third- to first-order Gauss-Hermite quadrature \citep{lb07}.  Table 1 of \cite{lb07} provides more details on the integration methods used as a function of $a$-$\vert u \vert$ space.  Our criteria for switching between the three computational methods is loosely based on \cite{lb07} and verified by testing and trial-and-error.

The second challenge is physical: the Lorentz, and hence the Voigt, profile over-estimates the far wings of the line profile due to pressure broadening (see \citealt{freedman08} and references therein). Even what ``far" actually means is not well understood.  Although this issue dominates the error budget, it is either treated as an ad hoc cutoff (in wavenumber) in the line wings (e.g., \citealt{sb07,amundsen14}), described qualitatively as a problem with no explicit cutoff being specified (e.g., \citealt{freedman08}) or simply left unmentioned (e.g., \citealt{irwin08,ms09,lee12,bs12,barstow13,lee13,line13}).  It is our hope that this issue will be acknowledged more explicitly and transparently in future studies involving atmospheric radiative transfer and retrieval.

In the current study, we do not attempt to solve this physics problem, which requires a detailed quantum mechanical calculation.  In the absence of a complete, first-principles theory, we instead compare calculations with the full Voigt profile versus those with some arbitrary line-wing cutoff specified, which we nominally take to be 500 Lorentz widths.  We emphasize that there is no sound physical reason behind choosing this particular cutoff.  It is merely used as a proof-of-concept comparison against calculations utilizing the full Voigt profile.

\subsection{GPU computing: memory types and parallelization}

We develop our custom-built code (\texttt{HELIOS-K}) using the native language of the NVIDIA GPUs, \texttt{CUDA} (Compute Unified Device Architecture), which is basically an embellished version of the \texttt{C} programming language \citep{sk10}.  A major advantage provided by a GPU is the large number of computational cores per card ($\sim 1000$) for a very low cost ($\sim \$1$ per core).  When compared head-to-head, a single GPU will always lose out against a single CPU in terms of both computational power and memory---the point is that one wins by throwing many, many more GPU cores at the problem.  A set of 32 consecutive threads is called a ``warp" and it is crucial that every warp performs exactly the same operation in order to optimise performance.  If not, a ``branch divergence" occurs and some operations are performed in serial operation mode.  Each calculation is performed on a thread and all of the threads are organised into blocks.

An indispensible part of writing ultrafast \texttt{CUDA} code is to understand the memory design and types on a GPU.  Global or device memory is the most abundant and can be accessed by every thread, but is generally the slowest type.  Shared memory is faster, but may only be accessed by threads within the same block.  Typically, a well-written \texttt{CUDA} kernel (usually called a ``function" in other languages) reads data from global into shared memory, performs the necessary arithmetic operations and writes back to global memory.  Another bottleneck is the passing of information (communication) between the CPU and GPU.  Exploiting the order-of-magnitude speed-ups a GPU has to offer is an exercise in shrewd memory and communication management.  Rather than describe each and every computing trick we used, we highlight the main ones and refer the reader to our open-source code.

For our application, we need to compute the Voigt profile for an enormous number of spectral lines across an even larger number of grid points in wavenumber.  Furthermore, we need to repeat this calculation for multiple combinations of temperature and pressure.  It is impossible to perform this computation in a single step, but we may perform a serial loop across the lines and parallelise across wavenumber.  This allows us to accumulate values of $\kappa(x)$ directly within a register (i.e., fastest available memory) without additional write-outs to global memory.

\subsection{Sorting and resampling}

Parallel sorting on a GPU is a non-trivial task.  Fortunately, this has already been implemented as part of the \texttt{CUDA} library (\texttt{https://developer.nvidia.com/Thrust}).  Once we have computed $\kappa(x)$, the challenge is to perform the sorting within each bin.  Each bin has a width $\Delta x$ and the number of bins typically used is $\sim 10$--$10^4$.  Sorting each bin in a serial fashion would be inefficient when the number of bins becomes large.  Instead, we sort the entire opacity function all at once, but keep track of the bin number each opacity point belongs to, which ultimately allows us to reconstruct $\kappa(y)$ in the individual bins.

Once we have sorted $\kappa(x)$ and obtained $\kappa(y)$, we wish to resample $\kappa(y)$ such that it is defined using a much smaller number of points (by orders of magnitude).  Numerous resampling strategies exist, including least-squares fitting, fast Fourier transforms, etc.  We find that using a least-squares fit with Chebyshev polynomials gives the best outcome in terms of accuracy and efficiency, especially since one may exploit the recurrence relations to generate Chebyshev polynomials of different orders.  We perform the fit on $\ln{\kappa(y)}$ to avoid numerical oscillations.  The least-squares fitting essentially involves solving $\hat{A} \vec{C} = \vec{D}$ for the vector of Chebyshev coefficients ($\vec{C}$), where $\vec{D}$ is the data vector.  Directly computing the inverse of the matrix $\hat{A}$ is expensive; instead, we implement ``Q-R decomposition" to obtain $\vec{C}$ \citep{nr}.  The final product of this step is a set of 20 Chebyshev coefficients describing $\kappa(y)$ for each bin.

\section{Results}

Unless otherwise stated, our results are based on computing a pure-water opacity function using the HITEMP line list, which consists of $\sim 10^8$ spectral lines of water.  We emphasize that this is a proof of concept and that \texttt{HELIOS-K} may be used for general mixtures of atoms and molecules.

\subsection{Basic setup}

We base the discussion of our results on a fiducial setup.  We focus on computing the opacity function for the water molecule, since it has the most lines among the major molecules expected (compared to CO and CO$_2$) and has the least controversial line list available (compared to CH$_4$).  In Figure \ref{fig:main}, we show two instances of the opacity function: one computed using the full Voigt profile and the other with a line-wing cutoff applied.  We divide the wavenumber region into three equal ranges: 0.5--8573.5 cm$^{-1}$ (infrared to near-infrared; $\gtrsim 1.2$ $\mu$m), 8573.5--17146.5 cm$^{-1}$ (near-infrared to optical; 0.6--1.2 $\mu$m) and 17146.5--25719.5 cm$^{-1}$ (optical; 0.4--0.6 $\mu$m).  Each bin has a width of $\Delta \nu = 8573$ cm$^{-1}$.  Within each bin, we adopt a resolution of $N_\nu = \Delta \nu / \delta \nu = 10^3$; we will demonstrate later that this attains convergence.  Our results point to the same qualitative conclusions even when more bins are used (not shown).

It is readily apparent that the choice of cutoff is a significant source of error in the near-infrared and optical, because it affects the weak lines more strongly, even prior to the mapping of the opacity function to its k-distribution counterpart.  We emphasize that this problem remains, regardless of whether the k-distribution method is used.  For the k-distribution function and the transmission function, the influence of the choice of line-wing cutoff is seen to be significant.  We will investigate this issue in more detail.

\subsection{Resampling as an insignificant source of error}

A necessary, intermediate step to check is whether the resampling of the k-distribution function using least-squares fitting introduces a significant source of error to our results.  In Figure \ref{fig:resampling}, we compute the transmission function in two ways: using the direct output from the mapping of $\kappa(x)$ to $\kappa(y)$ and the resampled $\kappa(y)$.  The difference between the two calculations is typically $\ll 1\%$ when $N_\nu \ge 10^3$.  Remarkably, resampling is not a significant source of error independent of the value of the column mass, i.e., it is equally robust in both optically thin and thick parts of the atmosphere.

\subsection{Choosing the correct bin resolution}

Even though convergence within each bin is tied to the number of lines present, we find that an easier rule of thumb is to use a minimum value of $N_\nu$ as a convergence criterion.  Figure \ref{fig:resampling} shows that convergence is comfortably attained for $N_\nu \ge 1000$.  This conclusion holds even when 1000 bins are used (not shown).

\subsection{Line-wing cutoff as the largest source of error}

\begin{figure}
\begin{center}
\vspace{-0.1in}
\includegraphics[width=\columnwidth]{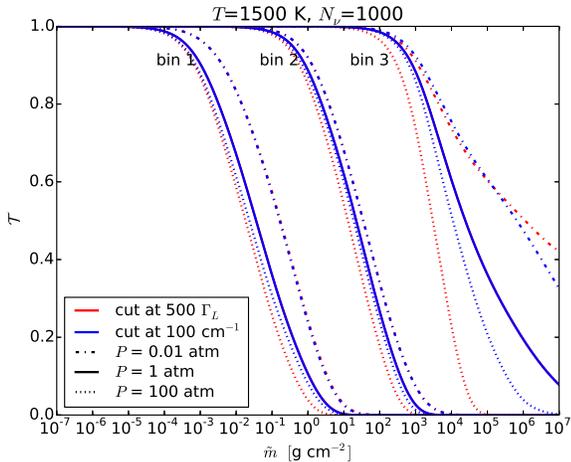}
\end{center}
\vspace{-0.2in}
\caption{Transmission function subjected to different choices of the line-wing cutoff and at different pressures.  The cuts of 100 cm$^{-1}$ and 500 $\Gamma_{\rm L}$ are chosen to match each other at $P=1$ atm.}
\label{fig:cutoff}
\end{figure}

\begin{figure}
\begin{center}
\vspace{-0.1in}
\includegraphics[width=\columnwidth]{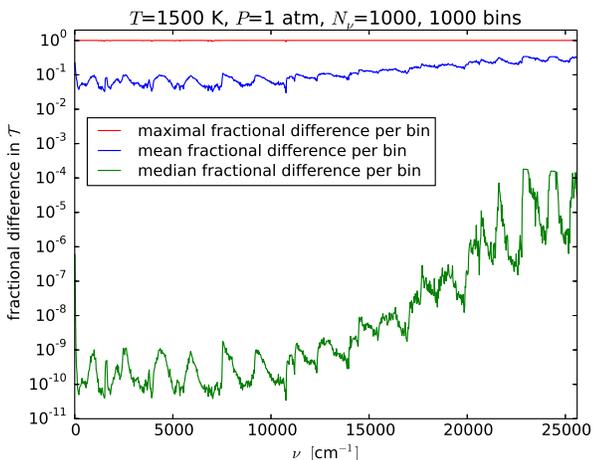}
\end{center}
\vspace{-0.2in}
\caption{Fractional difference in transmission function, between calculations using the full Voigt profiles and with a cutoff of $\Gamma_{\rm L}$ imposed, for 1000 bins across the same wavenumber range.}
\label{fig:1000bins}
\end{figure}

We further explore our claim that the line-wing cutoff is the largest source of error in computing and using an opacity function, regardless of whether one uses the k-distribution method.  In Figure \ref{fig:cutoff}, we show different calculations of ${\cal T}$ for various cutoff choices: an absolute cutoff (of 100 cm$^{-1}$, following the choice made by \citealt{sb07}) and an ad hoc cutoff of 500 Lorentz widths.  These choices are made such that they produce the same results at $P=1$ atm.  At higher pressures, we see that deviations appear.  For a given value of the column mass, the error is $\sim 10\%$ to even a factor of several in some instances.

We quantify this error in more detail.  Figure \ref{fig:1000bins} shows the fractional difference in ${\cal T}$, between calculations using the full Voigt profiles and those with a cutoff of $\Gamma_{\rm L}$ imposed, for 1000 bins across the same wavenumber range.  Across a broad range of column masses ($10^{-7} \le \tilde{m} \le 10^7$ g cm$^{-2}$), we compute the median, mean and maximum fractional differences using the full-Voigt calculations as a baseline comparison.  (We emphasize this does \textit{not} imply that using the full Voigt profile is correct.)  The median and maximum fractional differences are dominated by small and large column masses, respectively, and are not representative, but we show them for completeness.  We see that the mean fractional difference is $\sim 10\%$ across all wavenumbers for $P=1$ atm, implying that a similar uncertainty is present for the computed flux or synthetic spectrum.  We expect that the median fractional differences are larger for higher pressures.  Elucidating the full consequences of the uncertainty, associated with pressure broadening, for calculations of radiative transfer and retrieval is deferred to future work.

Generally, we find that the uncertainties associated with the line-wing cutoff are typically larger than those due to, e.g., resampling, as long as a sufficient bin resolution is used ($N_\nu \ge 10^3$, as previously demonstrated).

\subsection{Performance}

\begin{figure}
\begin{center}
\includegraphics[width=\columnwidth]{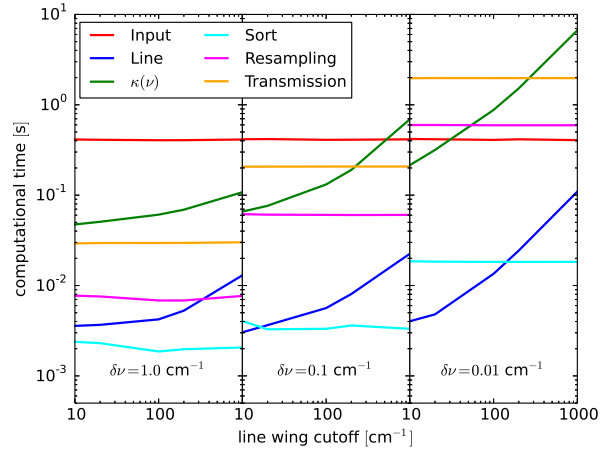}
\end{center}
\caption{Performance of \texttt{HELIOS-K} broken down into the various computational tasks: reading in line-list data (``Input"), computing Lorentz width, Doppler width, line strength and line center shift (``Line"), computing Voigt profiles and the opacity function (``$\kappa(\nu)$"), sorting values of $\kappa(x)$ into $\kappa(y)$ (``Sort"), resampling $\kappa(y)$ (``Resampling") and computing ${\cal T}$ for 1000 values of $\tilde{m}$ (``Transmission").  For this plot only, we are using the HITRAN database with $\sim 10^5$ water lines.}
\label{fig:performance}
\end{figure}

We execute performance tests on a NVIDIA Tesla K20 GPU card, which has 2496 cores.  For these tests, we use the HITRAN ($\sim 10^5$ water lines), instead of the HITEMP, line list, as the entire calculation fits within a single K20 GPU card.  (The HITEMP water line list is provided in 34 separate chunks, which we simply load in serial.)  Figure \ref{fig:performance} breaks down the performance of our code, which we name \texttt{HELIOS-K}, in terms of the various tasks executed.  Unsurprisingly, the computational cost goes up with bin resolution and line-wing cutoff.  Generally, \texttt{HELIOS-K} takes $\sim 1$ s to compute $\sim 10^5$ spectral lines of water.  We anticipate that such a level of performance allows for efficient and broad sweeps of the parameter space of exoplanetary atmospheres.

\section{Discussion \& Conclusions}

\subsection{Towards uniform standards: a checklist for opacity function calculations}

The details of how opacity functions are computed and used by various studies in the literature remain vague or incomplete.  We suggest that a path towards uniform standards involves explicitly addressing the following questions (and publishing the answers to them).
\begin{itemize}

\item Does the study claim a ``line-by-line" calculation of the opacity function (e.g., \citealt{ms09,bs12})?  If so, are the lines being sampled in an adequate way?  E.g., if there are $N_{\rm lines}$ lines, is $N_{\rm sample} \gg N_{\rm lines}$, where $N_{\rm sample}$ is the number of wavenumber/wavelength points used?  If special circumstances (e.g., very broad lines) allow for $N_{\rm sample} \sim N_{\rm lines}$ to be justified, has this been demonstrated explicitly?  Does the study show results from convergence tests?  Often, what are effectively opacity-sampling techniques ($N_{\rm sample} \ll N_{\rm lines}$) are misleadingly claimed as being ``line-by-line".

\item How is the Voigt profile being computed?  Is it being directly evaluated as an indefinite integral?  Or has a transformation and/or approximation(s) been taken?

\item If the k-distribution method is adopted, how many bins are specified?  How is the opacity function resampled within each bin, i.e., what is the resampling method?  Has the study demonstrated that an adequate intra-bin resolution has been used?

\item Are k-distribution tables separately computed for each molecular species and then added together---weighted by the relative abundance of each species---afterwards?  If so, then the correlated-k approximation has been used and this should be explicitly mentioned.

\item Is pressure broadening being considered?  If so, is the study imposing line-wing cutoffs?  Is the cutoff specified as an absolute number or as a specific number of Lorentz or Doppler widths?  Have the uncertainties associated with this choice been explored and quantified?

\end{itemize}

The preceding checklist may be a useful guide for reviewing studies that perform radiative transfer or retrieval calculations.

\subsection{Summary}

We have constructed an open-source, ultrafast, GPU code written using \texttt{CUDA}, named \texttt{HELIOS-K}, which takes a line list as an input and computes the opacity function of the atmosphere for any mixture of atoms and molecules.  The dominant source of error stems from an unsolved physics problem: describing the far line wings of spectral lines affected by pressure broadening.  In the absence of a complete theory, we (and others before us) have applied an ad hoc cutoff of the line wing for our calculations.  Notwithstanding this issue, \texttt{HELIOS-K} provides the exoplanet community with an efficient tool for computing opacity functions.

\acknowledgments
S.L.G. and K.H. acknowledge support from the Swiss National Science Foundation and the Swiss-based MERAC Foundation for grants awarded to the Exoclimes Simulation Platform (\texttt{www.exoclime.org}).  We thank Chris Hirata for illuminating conversations and the Exoplanets \& Exoclimes Group for interactions and intellectual stimulation.  We are grateful to David Amundsen, Isabelle Baraffe and the anonymous referee for constructive comments that improved the clarity of the manuscript.


\label{lastpage}


\begin{thebibliography}{99}

\bibitem[Amundsen et al.(2014)]{amundsen14} Amundsen, D.S., Baraffe, I., Tremblin, P., Manners, J., Hayek, W., Mayne, N.J., \& Acreman, D.M. \ 2014, A\&A, 564, A59

\bibitem[Barstow et al.(2013)]{barstow13} Barstow, J.K., Aigrain, S., Irwin, P.G.J., Bowles, N., Fletcher, L.N., \& Lee, J.-M. \ 2013, MNRAS, 430, 1188

\bibitem[Benneke \& Seager(2012)]{bs12} Benneke, B., \& Seager, S. \ 2012, ApJ, 753, 100

\bibitem[Brown(2001)]{brown01} Brown, T.M. \ 2001, ApJ, 553, 1006

\bibitem[Burrows et al.(1997)]{burrows97} Burrows, A., et al. \ 1997, ApJ, 491, 856

\bibitem[Burrows et al.(2001)]{burrows01} Burrows, A., Hubbard, W.B., Lunine, J.I., \& Liebert, J. \ 2001, Reviews of Modern Physics, 73, 719

\bibitem[Charbonneau(2009)]{char09} Charbonneau, D. \ 2009, Proceedings of the International Astronomical Union, 253, 1

\bibitem[Draine(2011)]{draine11} Draine, B.T. \ 2011, Physics of the Interstellar and Intergalactic Medium (New Jersey: Princeton University Press)

\bibitem[Fortney et al.(2010)]{fortney10} Fortney, J.J., Shabram, M., Showman, A.P., Lian, Y., Freedman, R.S., Marley, M.S., \& Lewis, N.K. \ 2010, ApJ, 709, 1396

\bibitem[Freedman, Marley \& Lodders(2008)]{freedman08} Freedman, R.S., Marley, M.S., \& Lodders, K. \ 2008, ApJS, 174, 504

\bibitem[Fu \& Liou(1992)]{fu92} Fu, Q., \& Liou, K.N. \ 1992, Journal of the Atmospheric Sciences, 49, 2139

\bibitem[Goody \& Yung(1989)]{gy89} Goody, R.M., \& Yung, Y.L. \ 1989, Atmospheric Radiation: Theoretical Basis, 2nd edition (New York: Oxford University Press)

\bibitem[Heng, Mendon\c{c}a \& Lee(2014)]{hml14} Heng, K., Mendon\c{c}a, J.M., \& Lee, J.-M. \ 2014, ApJS, 215, 4

\bibitem[Heng \& Showman(2015)]{hs15} Heng, K., \& Showman, A.P. \ 2015, AREPS, in press (arXiv:1407.4150)

\bibitem[Irwin et al.(2008)]{irwin08} Irwin, P.G.J., et al. \ 2008, JQSRT, 109, 1136

\bibitem[Lacis \& Oinas(1991)]{lo91} Lacis, A.A., \& Oinas, V. \ 1991, Journal of Geophysical Research, 96, 9027

\bibitem[Lee, Fletcher \& Irwin(2012)]{lee12} Lee, J.-M., Fletcher, L.N., \& Irwin, P.G.J. \ 2012, MNRAS, 420, 170

\bibitem[Lee, Heng \& Irwin(2013)]{lee13} Lee, J.-M., Heng, K., \& Irwin, P.G.J. \ 2013, ApJ, 778, 97

\bibitem[Letchworth \& Benner(2007)]{lb07} Letchworth, K.L., \& Benner, D.C. \ 2007, JQSRT, 107, 173

\bibitem[Line et al.(2013)]{line13} Line, M.R., et al. \ 2013, ApJ, 775, 137

\bibitem[Madhusudhan \& Seager(2009)]{ms09} Madhusudhan, N., \& Seager, S. \ 2009, ApJ, 707, 24 

\bibitem[Madhusudhan et al.(2014)]{madhu14} Madhusudhan, N., Knutson, H., Fortney, J.J., \& Barman, T. \ 2014, in Protostars \& Planets IV, eds. H. Beuther, R.S. Klessen, C.P. Dullemond and T. Henning, 739--762 (Tucson: University of Arizona Press)

\bibitem[Marley et al.(1996)]{marley96} Marley, M.S., Saumon, D., Guillot, T., Freedman, R.S., Hubbard, W.B., Burrows, A., \& Lunine, J.I. \ 1996, Science, 272, 1919

\bibitem[Mihalas(1970)]{mihalas} Mihalas, D. \ 1970, Stellar Atmospheres (San Francisco: Freeman)

\bibitem[Penner(1952)]{penner52} Penner, S.S. \ 1952, Journal of Chemical Physics, 20, 507

\bibitem[Rothman et al.(1996)]{rothman96} Rothman, L.S., et al. \ 1996, Journal of Quantitative Spectroscopy \& Radiative Transfer, 60, 665

\bibitem[Rothman et al.(2010)]{rothman10} Rothman, L.S., et al. \ 2010, Journal of Quantitative Spectroscopy \& Radiative Transfer, 111, 2139

\bibitem[Rothman et al.(2013)]{rothman13} Rothman, L.S., et al. \ 2013, Journal of Quantitative Spectroscopy \& Radiative Transfer, 130, 4

\bibitem[Pierrehumbert(2010)]{pierrehumbert} Pierrehumbert, R.T. \ 2010, Principles of Planetary Climate (New York: Cambridge University Press)

\bibitem[Press et al.(2007)]{nr} Press, W.H., Teukolsky, S.A., Vetterling, W.T., \& Flannery, B.P. \ 2007, Numerical Recipes: The Art of Scientific Computing, third edition (New York: Cambridge University Press)

\bibitem[Sanders \& Kandrot(2010)]{sk10} Sanders, J., \& Kandrot, E. \ 2010, CUDA by Example: An Introduction to General-Purpose GPU Programming (Indianapolis: Addison-Wesley)

\bibitem[Seager \& Deming(2010)]{sd10} Seager, S., \& Deming, D. \ 2010, ARA\&A, 48, 631

\bibitem[Sharp \& Burrows(2007)]{sb07} Sharp, C.M., \& Burrows, A. \ 2007, ApJS, 168, 140

\bibitem[Showman et al.(2009)]{showman09} Showman, A.P., Fortney, J.J., Lian, Y., Marley, M.S., Freedman, R.S., Knutson, H.A., \& Charbonneau, D. \ 2009, Astrophysical Journal, 699, 564

\bibitem[Zaghloul \& Ali(2012)]{za12} Zaghloul, M.R., \& Ali, A.N. \ 2012, ACM Transactions on Mathematical Software, 38, 15 (arXiv:1106.0151)

\end{thebibliography}
\end{document}